\documentclass[aip,jcp,showpacs,preprint]{revtex4-1}

\usepackage[version=3]{mhchem}
\usepackage{graphicx}
\usepackage{pstricks}
\usepackage{amssymb}
\usepackage{amsmath}
\usepackage{amsfonts}

\newcommand{\beq}{\begin{equation}}
\newcommand{\eeq}{\end{equation}}
\newcommand{\bea}{\begin{eqnarray}}
\newcommand{\eea}{\end{eqnarray}}

\begin{document}
%
%

\title {Non-invasive estimation of dissipation \\ from non-equilibrium fluctuations in chemical reactions}
\author{S. Muy}
\affiliation{Laboratoire de Physico-Chimie Th\'eorique - UMR CNRS Gulliver 7083,
ESPCI, 10 rue Vauquelin, F-75231 Paris, France}
\author{A. Kundu}
\affiliation{Laboratoire de Physique Th\'eorique et Mod\`eles Statistiques - UMR CNRS 8626,
Universit\'e Paris-Sud, B\^at. 100, 91405 Orsay Cedex, France}
\author{D. Lacoste$^1$}

\date{\today}

\begin{abstract}
We show how to extract from a sufficiently long time series of
stationary fluctuations of chemical reactions
 an estimate of the entropy production.
This method, which is based on recent work on fluctuation theorems,
is direct, non-invasive, does not require any knowledge about
 the underlying dynamics, and is applicable
 even when only partial information is available. We apply it to
 simple stochastic models of chemical reactions involving
 a finite number of states,
 and for this case, we study how the estimate of dissipation is affected
 by the degree of coarse-graining present in the input data.
\end{abstract}

\maketitle

\section{Introduction}
Stochastic fluctuations play a key role in many living processes, in particular at
the molecular level in cells, where many complex processes occur (non-regular
feedback, regulation, proof-reading..) on different time scales. These fluctuations are intrinsically non-equilibrium in nature. With the development
of single molecule techniques and
chemical sensors, more and more data representing non-equilibrium
fluctuations of small objects is becoming available in soft matter and
biology. With all these developments, the question of extracting relevant
information from an increasing amount of experimental data of this kind is
becoming central.

One example of relevant information is whether the fluctuations
originate from active, energy consuming, processes
or from passive, equilibrium like, processes.
Ideally, one would like to distinguish one from the other
directly from trajectory information, without any knowledge
 of the dynamics, and one
would also like to measure the distance from equilibrium by
an estimate of the dissipation or entropy production.
One possibility to do this is to first infer the rate constants entering in the dynamics,
by studying for instance the response of the system to a perturbation \cite{Berthoumieux2009} or
by using Bayesian inference techniques. One can then test directly whether detailed balance holds.
Alternatively, one can also determine whether the fluctuation-dissipation theorem (FDT) holds without
 trying to infer the dynamics first \cite{Mizuno2007_vol315}. Although these are useful methods,
there are also several limitations to such approaches: (i) the system
must be perturbed and (ii) the determination of the entropy
production from the violation of the FDT is not straightforward in general.
The latter requires a determination of the dynamics which furthermore needs to be of the Langevin
type \cite{Harada2005_vol95,Lander2012}. Such an approach is 
related to many recent studies on a modified fluctuation-dissipation theorem 
near non-equilibrium steady states \cite{Verley2011a,Seifert2010_vol89,Chetrite2008_vol2008}. 

In the present paper, we apply a recent method \cite{Roldan2010_vol105} to
estimate the dissipation from trajectory
information only. This method avoids the drawbacks (i) because it does not require to apply a
perturbation, it is non-invasive, and (ii) because it gives direct access to the
entropy production. 
The method is based on a connection between two measures of irreversibility \cite{Kawai2007_vol98},
similar to the Landauer principle linking dissipation and information processing
and is rooted in recent progresses on fluctuation theorems.
The first measure of irreversibility is characterized by the thermodynamic
 notion of entropy production. The second measure
corresponds to the temporal asymmetry of the fluctuations \cite{Maes2003_vol110},
 it is an information theoretic quantity constructed from the trajectories.
This temporal asymmetry is related to the difference between the
dynamical randomness associated with a direct or forward path and that associated with an
appropriate reverse path, in which the driving must be reversed \cite{Gaspard2004_vol117a}.
In the particular case of non-equilibrium steady-states, a simpler formulation is
possible because the driving, if present, is independent of time,
and therefore there is no need to reverse the driving in the backward process \cite{Roldan2010_vol105}.

In section II below, we present the general principle of the method to estimate the dissipation using non-equilibrium fluctuations. The method is then illustrated using a linear three state enzymatic network. 
We note at the end of this section that the method requires 
a knowledge of all the chemical transitions which occur at a given time. An experiment will hardly give access to such a detailed information. Checking how the evaluation of the entropy production 
is affected by the partial knowledge of the system is therefore necessary and this will be done in section III, 
using the same three state model as example. In section IV, we show that the method is not limited to linear models 
but is applicable to non-linear models such as 
the Schl\"ogl \cite{Schlogl1972} and the Schnakenberg models \cite{Schnakenberg1979}.

\section{Stochastic description of entropy production from non-equilibrium trajectories}
In this approach, one considers a block $x_1..x_m$ of length $m$ of the
 original stationary series assumed to be
much longer, of length $n \gg m$. We assume that the random variables $x_1..x_m$
can only take discrete values. Let us denote by $p_F=p(x_1..x_m)$ the probability
 to observe that block when reading
the series forward in time, while $p_B=p(x_m..x_1)$ represents the probability to observe
 the time-reversed block under the same conditions.
A key quantity is the relative entropy between these two distributions:
\beq
D_m(p_F|p_B)=\sum_{x_1..x_m} p(x_1..x_m) \ln \frac{p(x_1..x_m)}{p(x_m..x_1)}.
\label{rel entropy}
\eeq
In the case that $p_F$ and $p_B$ contain the full information about the dynamics
 (in a sense to be made more precise below), one has the following {\it equality}
 in the space of trajectories, with the Boltzmann constant set to unity \cite{Horowitz2009}:
\beq
\langle \Delta S \rangle = d(p_F | p_B)=\lim_{m \rightarrow \infty} \frac{1}{m} D_m(p_F|p_B),
\label{main result}
\eeq
where $\langle \Delta S \rangle$ represents the mean
entropy production rate of the trajectory. Here, the notation $\langle .. \rangle$ means a statistical average with respect to the probability distribution $p(x_1..x_m)$, which in the present case is evaluated using a single very long trajectory. 
In practice, instead of evaluating the right hand side of this equation, one can
obtain $d(p_F | p_B)$ as the limit of $d_m= D_m-D_{m-1}$ for $m \rightarrow \infty$,
which shows a faster convergence \cite{Schuermann1996}.
For Markovian dynamics, one can show that for any $m \ge 0$, $d(p_F | p_B)=d_{m+2}=d_2$.
Numerically, $d(p_F | p_B)$ is detected as a plateau for large $n$ when varying the length of the trajectory $n$.

\subsection{Application to a three-state enzymatic network}
In this paper, we apply this method to chemical reactions ruled by master equation,
thus going beyond the examples of Ref.~\cite{Roldan2010_vol105} which involved only one degree of freedom.
As a paradigm for such networks, we consider the following three-state enzymatic network \cite{Ge2012,Berthoumieux2009}.
\beq
\ce{A <=>[k_1][k_{-1}] B}, \,\,\,\, \ce{B <=>[k_2][k_{-2}] C}, \,\,\,\, \ce{C <=>[k_3][k_{-3}] A},
\eeq
where $k_i$ denote the rate constants. We consider a pool of $N$ particles of this kind.
A state of this system can be represented by $c=(n_A,n_B)$, where $n_A$ and $n_B$ are particle numbers
of species $A$ and $B$ given the existence of
 the conservation law $N=n_A+n_B+n_C$.

Let us call $p_t(n_A,n_B,n_C)$ the probability to observe
such a state at the time $t$. It obeys the following chemical master equation:
\bea
\frac{d p_t(n_A,n_B,n_C)}{dt} &=& k_1 (n_A +1) p_t(n_A+1,n_B-1,n_C)+k_{-3} (n_A+1) p_t(n_A+1,n_B,n_C-1) \nonumber \\
&+& k_2 (n_B+1) p_t(n_A,n_B+1,n_C-1)+k_{-1} (n_B +1) p_t(n_A-1,n_B+1,n_C) \nonumber \\
&+& k_3 (n_C +1) p_t(n_A-1,n_B,n_C+1) + k_{-2} (n_C +1) p_t(n_A,n_B-1,n_C+1) \nonumber \\
&-& (k_1 n_A + k_2 n_B + k_3 n_C + k_{-1} n_B + k_{-2} n_C +k_{-3} n_A) p_t(n_A,n_B,n_C),
\eea
which is to be solved with the conservation law $N=n_A+n_B+n_C$.
The solution of this equation has the form of a multinomial distribution \cite{Hill1965,Hill1971}:
\beq
p_t(n_A,n_B,n_C)= \frac{N!}{n_A ! \, n_B ! \, n_C !} p_A(t)^{n_A} p_B(t)^{n_B} p_B(t)^{n_B},
\label{Hill}
\eeq
where $p_A(t)$ (respectively $p_B(t)$ and $p_C(t)$) is the probability of a given particle to belonging to the chemical species $A$ (respectively $B$ and $C$) at time $t$. This
results holds at any time. In the stationary state, this solution is denoted by $p_{st}(n_A,n_B,n_C)$.

From Eq.~\ref{Hill}, one deduces that for any $i \in (A,B,C)$, the mean of the random variable $n_i$ is $\langle n_i \rangle =N p_i$ and the variance is $\langle (n_i-\langle n_i \rangle)^2 \rangle=N p_i (1-p_i)$.
 As a result, in a volume $\Omega$, the concentration of the chemical species $i$, denoted by $[i]$, 
is equal to $p_i$ multiplied by the
 total concentration $N/\Omega$, \textit{i.e.} $[i]=\langle n_i \rangle /\Omega=Np_i/\Omega$.
In the limit of $\Omega \rightarrow
\infty$, the concentrations and the variables $p_i$ obey the deterministic rate equations of chemical kinetics. 
In the present case, these rate equations are:
\bea
\frac{d p_A}{dt} &=& k_3 p_C + k_{-1} p_B - (k_{-3} + k_1 ) p_A \nonumber \\
\frac{d p_B}{dt} &=& k_1 p_1 + k_{-2} p_C - (k_{-1} + k_2 ) p_B \nonumber \\
\frac{d p_C}{dt} &=& k_{-3} p_A + k_2 p_B - (k_{-2} + k_3 ) p_A.
\label{linear system}
\eea


It is also a simple calculation to show that the average entropy production rate (EPR) is, 
\beq
\langle \Delta \dot{S} \rangle=N \frac{k_1 k_2 k_3 - k_{-1} k_{-2} k_{-3}}{K} \ln \frac{k_1 k_2 k_3}{k_{-1} k_{-2} k_{-3}},
\label{DS linear pb}
\eeq
where $K$ is the constant
\beq
K=k_1 k_3 + k_{-1} k_{-2} + k_{-1} k_{-3} + k_{-1} k_3 + k_2 k_1 + k_2 k_{-3} + k_2 k_3 +k_{-2} k_1 + k_{-2} k_{-3}.
\eeq

From this expression, it follows that the condition for which this system
reaches a non-equilibrium steady state different from equilibrium is $k_1 k_2 k_3 \neq k_{-1} k_{-2} k_{-3}$.
This result holds in fact for any $N$ due to the linearity of the equation with respect to $N$.
Note also that this entropy production rate takes the form of a sum of products of generalized forces and fluxes.
Here the flux is the stationary current
$J_{st}=N (k_1 k_2 k_3 - k_{-1} k_{-2} k_{-3})/K$, while the thermodynamic force is (in units of $k_BT$ following the convention of Ref.~ \cite{Callen1985})
$A=\ln k_1 k_2 k_3/k_{-1} k_{-2} k_{-3}$, the affinity of the cycle $A \rightarrow B \rightarrow C \rightarrow A$.
More generally, for any non-equilibrium steady state (NESS), the mean entropy production rate can be expressed as a sum of
product of fluxes and affinities on the cycles of a fundamental set \cite{Schnakenberg1976}.

We are now in position to explain how to recover Eq.~\ref{DS linear pb} using the formulation
of Eqs.~\ref{rel entropy}-\ref{main result}. To do so, we have simulated numerically the chemical master equation
using the Gillespie algorithm \cite{Gillespie1977_vol173},
which generates the correct exponential distribution of waiting times between two consecutive events. 
We have then used the time series constructed in this way to evaluate 
the dissipation using Eqs.~\ref{rel entropy}-\ref{main result}.
Since Eq.~\ref{main result} represents an entropy per data in discrete time while the Gillespie 
algorithm is formulated in continuous time, we introduce 
the characteristic time per data $\tau$ defined below to convert the discrete time formulation to the continuous one. 

Let us consider a Markovian dynamics so that we only need to focus on $D_2$. 
According to Eq.~\ref{rel entropy}, $D_2$ is
\beq
D_2=\sum_{c,c'} p(c,c') \ln \frac{p(c,c')}{p(c',c)},
\label{D_2}
\eeq
where $c$ and $c'$ refer to two consecutive values of the state vector $(n_A,n_B)$
in the time series which is analyzed. In this case, the expression in Eq.~\ref{D_2} becomes
\beq
D_2=\sum_{c,c'} p(c,c') \ln \frac{p(c | c')}{p(c' | c)},
\eeq
where $p(c|c')$ represents the conditional probability to go from state $c'$ to $c$.
In a Markov process, one can show that this conditional probability is related to
the transition rate by the expression
\beq
p(c | c')=\frac{w(c,c')}{\lambda(c')},
\eeq
where $w(c,c')$ is the transition rate to go from state $c'$ to $c$ and $\lambda(c)=\sum_{c' \neq c} w(c',c)$ represents the escape rate to leave state $c$.
It follows from the above equations that
\beq
D_2=\sum_{c,c'} \frac{w(c,c') p_{st}(c')}{\lambda(c')} \ln \frac{w(c,c') \lambda(c)}{w(c',c) \lambda(c')}.
\eeq
The average escape rate in this problem is
\beq
\tau=\sum_{c} p_{st}(c) \frac{1}{\lambda(c)}.
\eeq
At a mean-field level, for the three state enzymatic model, this time is 
\beq
\tau=\frac{1}{N \left( k_1 p_A + k_2 p_B + k_3 p_C + k_{-1} p_B + k_{-2} p_C + k_{-3} p_A \right)},
\label{explicit tau}
\eeq
where $p_A$, $p_B$ and $p_C$ are the stationary probability distributions introduced earlier.
We have verified that this expression provides a good estimate of the characteristic jump time, by comparing it
 with the mean duration between two configuration changes observed in the sequence, which can be
determined numerically. 
For more general situations, where a formula like
Eq.~\ref{explicit tau} is not be available,
the numerical determination is the only option.
The EPR is then obtained from $D_2$ and $\tau$ as,
\beq
\langle \Delta \dot{S} \rangle \simeq \frac{D_2}{\tau},
\eeq
where $\tau$ is approximated by the total time of the Gillespie simulation divided by the total
number of configuration changes or 'jumps'.

In figure \ref{S1}, we compare the numerical estimation of the
 EPR based on $D_2$ with the exact value given by Eq.~\ref{DS linear pb}. To do this we plot $D_2$ as function
of the trajectory length $n$ and we look for a plateau at large $n$.
As shown in figure \ref{S1}, we indeed find a plateau at the expected value of the EPR (solid line).
In fact, it is remarkable that we are able to recover in this way for any system size $N$,
 the expected exact value of the entropy production rate.
Varying $N$ at fixed rate constants only affects the characteristic time $\tau$ but not the relative entropy, as expected
 since time does not enter in Eq.~\ref{rel entropy}.

\section{Role of coarse-graining of the description in the estimation of entropy production}
In the three states enzymatic network studied above, we are 
able to recover the known amount of dissipation in this NESS
 uniquely from trajectory information without any knowledge of the underlying dynamics.
It is important to point out however, that this determination requires a knowledge of all the elementary
 transitions with a single molecule resolution.
For applications, such a time and particle number resolution will be hard to achieve,
which is why it is important to determine how the estimate of entropy production
is affected by the unavoidable coarse-graining of the original data.
Given that our method is based on a fluctuation theorem, a related question is
how coarse-graining affects fluctuation theorems \cite{Rahav2007,Kawaguchi2012}.
In general, Eq.~\ref{rel entropy} should hold as an inequality \cite{Gomez-Marin2008_vol78}, namely
\beq
\langle \Delta S \rangle \ge d(\tilde{p}_F | \tilde{p}_B)=\lim_{m \rightarrow \infty} \frac{1}{m}
D_m(\tilde{p}_F|\tilde{p}_B),
\label{main result cg}
\eeq
where $\tilde{p}$ represents a coarse-grained version of the path probability denoted $p$ above.

We provide below an illustration of this idea by distinguishing three forms of coarse-graining for the model introduced above:
(a) we discard one chemical specie, so we are given only the trajectory of the chemical species $A$, namely $n_A(t)$,
(b) we only have access to finite resolution in particle number or concentration, and (c) we only have access to finite 
resolution in time.
Note that coarse-graining due to decimation of fast states is included in (a) and (b); for instance, 
a coarse-graining of type (a) is considered in Ref.~\cite{Amann2010},
while a coarse-graining of type (b) is considered in Ref.~\cite{Puglisi2010_vol2010}.

Let us first consider the case (a), where only the variable $n_A$ is observed among the three variables $n_A$,$n_B$ and $n_C$.
In this situation, we show explicitly below that $D_2=0$.
According to Eq.~1, $D_2$ is given by
\beq
D_2=\sum_{n_A^1,n_A^2} p(n_A^1,n_A^2) \ln \frac{p(n_A^1,n_A^2)}{p(n_A^2,n_A^1)},
\eeq
where $n_A^1$ and $n_A^2$ refer to two consecutive values of $n_A$ in the time series which is analyzed. Since the case where $n_A^1=n_A^2$
clearly does not contribute to $D_2$ and we consider elementary reactions, we only need 
to consider $n_A^2=n_A^1 \pm 1$. After renaming $n_A^1$ by $n_A$, we obtain
\beq
D_2=\sum_{n_A=0}^{N-1} p(n_A,n_A+1) \ln \frac{p(n_A,n_A+1)}{p(n_A+1,n_A)}+\sum_{n_A=1}^{N} p(n_A,n_A-1) \ln \frac{p(n_A,n_A-1)}{p(n_A-1,n_A)},
\eeq
which can be rewritten as
\beq
D_2=\sum_{n_A=1}^N [ p(n_A - 1,n_A) - p(n_A,n_A - 1) ] \ln \frac{p(n_A -1, n_A)}{p(n_A,n_A - 1)}.
\label{D2}
\eeq
Note that $D_2$ depends on the quantity $J(n_A -1 \rightarrow n_A)=  p(n_A - 1,n_A) - p(n_A,n_A - 1)$,
which has the interpretation of the local current between $n_A-1$ and $n_A$ in discrete time. By going to
the level of description with the full information $(n_A,n_B,n_C)$ for which the evolution is Markovian, we obtain
\bea
& & p(n_A - 1,n_A) - p(n_A,n_A - 1) = \sum_{n_B,n_C} \left[ p(\{ n_A-1,n_B+1,n_C \}, \{n_A,n_B,n_C \}) \right. \nonumber \\
&+ &  p( \{n_A-1,n_B,n_C+1 \}, \{n_A,n_B,n_C \}) -  p( \{n_A,n_B,n_C \}, \{ n_A-1,n_B+1,n_C \}) \nonumber \\
&-& \left.
p(\{n_A,n_B,n_C \}, \{n_A-1,n_B,n_C+1 \}) \right], \nonumber
\eea
where it is understood that the sum over $n_B$ and $n_C$ is done with the condition $n_B+n_C=N-n_A$.
This sum can be written as
\beq
p(n_A - 1,n_A) - p(n_A,n_A - 1)=  \frac{n_A}{p_A} \sum_{n_B,n_C} \left( p_B k_{-1} + p_C k_3 - p_A(k_1 + k_{-3}) \right) p_{st}(n_A,n_B,n_C).
\eeq
In view of Eq.~\ref{linear system}, the term in the parenthesis is zero in the stationary state,
therefore $D_2=0$ for this model with partial information.
Alternatively, one can also show that $D_2=0$ from the following argument:
In a steady state in 1D, the local current $J$ must be global. It is then equal 
to $J= \lim_{t \rightarrow \infty} \langle n_A(t) \rangle/t$,
which is zero because $n_A$ is stationary.
Therefore $D_2$ is zero as shown in the inset of fig.~\ref{fig1}.

\begin{figure}[t!]
\includegraphics[scale=0.5]{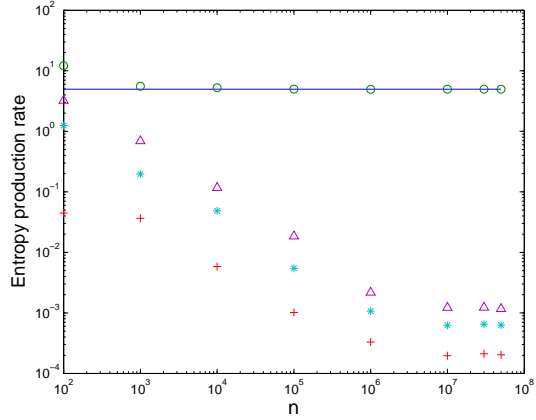}
\caption{\label{S1}
 Estimate of the entropy production rate (in units of $k_BT$ per $s$) as function of the
length of the trajectory $n$ using full
information with $D_2$ (circles) compared with the exact value of entropy production given by Eq.~\ref{DS linear pb} (solid line).
The other symbols correspond to estimates using partial information:
$D_3$ (plus), $D_4$ (stars) and $D_5$ (triangles). All the estimates are
divided by the characteristic time $\tau$, which is numerically
estimated from the trajectories.
The parameters are $k_1=1.2, k_{-1}=0.3, k_2=0.9, k_{-2}=0.4, k_{3}=0.5$, and $k_{-3}=0.2$.}
\end{figure}

\begin{figure}[t!]
\includegraphics[scale=0.5]{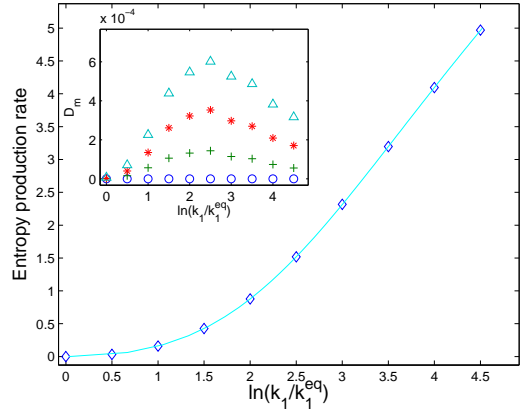}
\hspace{2mm}
\caption{\label{fig1}
 Entropy production rate (in units $s^{-1}$) as function of $\ln (k_1/k_1^{eq})$, as estimated
from $D_2$ (diamonds) using full information together with the analytical expression from Eq.~\ref{DS linear pb} (solid line).
The total number of particles is $N=5$, the length of trajectory is $n=4 \cdot 10^7$, and the rates
(in units $s^{-1}$) are $k_{-1}=0.3, k_2=0.9, k_{-2}=0.4, k_{3}=0.5$, and $k_{-3}=0.2$.
In the inset, $D_2$ (circles), which is zero, $D_3$ (plus), $D_4$ (stars) and $D_5$ (triangles) are shown as
function of the same log-ratio when only partial trajectory information is used.}
\end{figure}

In the case (a), the system appears stationary and the distribution of $n_A$ is
 given by the classical binomial distribution:
\beq
p(n_A)= \left( {\begin{array}{*{20}c} N \\ n_A \\ \end{array}} \right) p_A^{n_A} (1-p_A)^{N-n_A}.
\eeq
Thus, there is no possibility
 to suspect the existence of a current towards $B$ or $C$ or between $B$ and $C$ since we know nothing
 of these chemical species, in fact from a chemistry point of view, all evidences point towards equilibrium.
 And, yet the system is out of equilibrium, and remarkably this method can detect it.
Since the coarse-grained dynamics is non-Markovian in this case, $d_3, d_4, d_5..$ obtained from the 
trajectory of the ${n_A(t)}$ trajectory only, are non-zero and different from each other, because they contain information
 about higher correlations of the time series. This reveals that the original system is not in equilibrium.
We show in figure \ref{S1} how the various estimates of the entropy production vary as the length
of the trajectory $n$ increases. As mentioned in the previous section,
when full information is available, we obtain for $D_2$ a plateau
corresponding to the expected value of the entropy production rate calculated using Eq.~\ref{DS linear pb} (solid line).
In the case of partial information, there is also a convergence of $D_3$, $D_4$ and $D_5$  
as function of $n$ but the convergence is slower than for $D_2$ (the plateau occurs at larger values of $n$).
\begin{figure}[t!]
\includegraphics[scale=0.5]{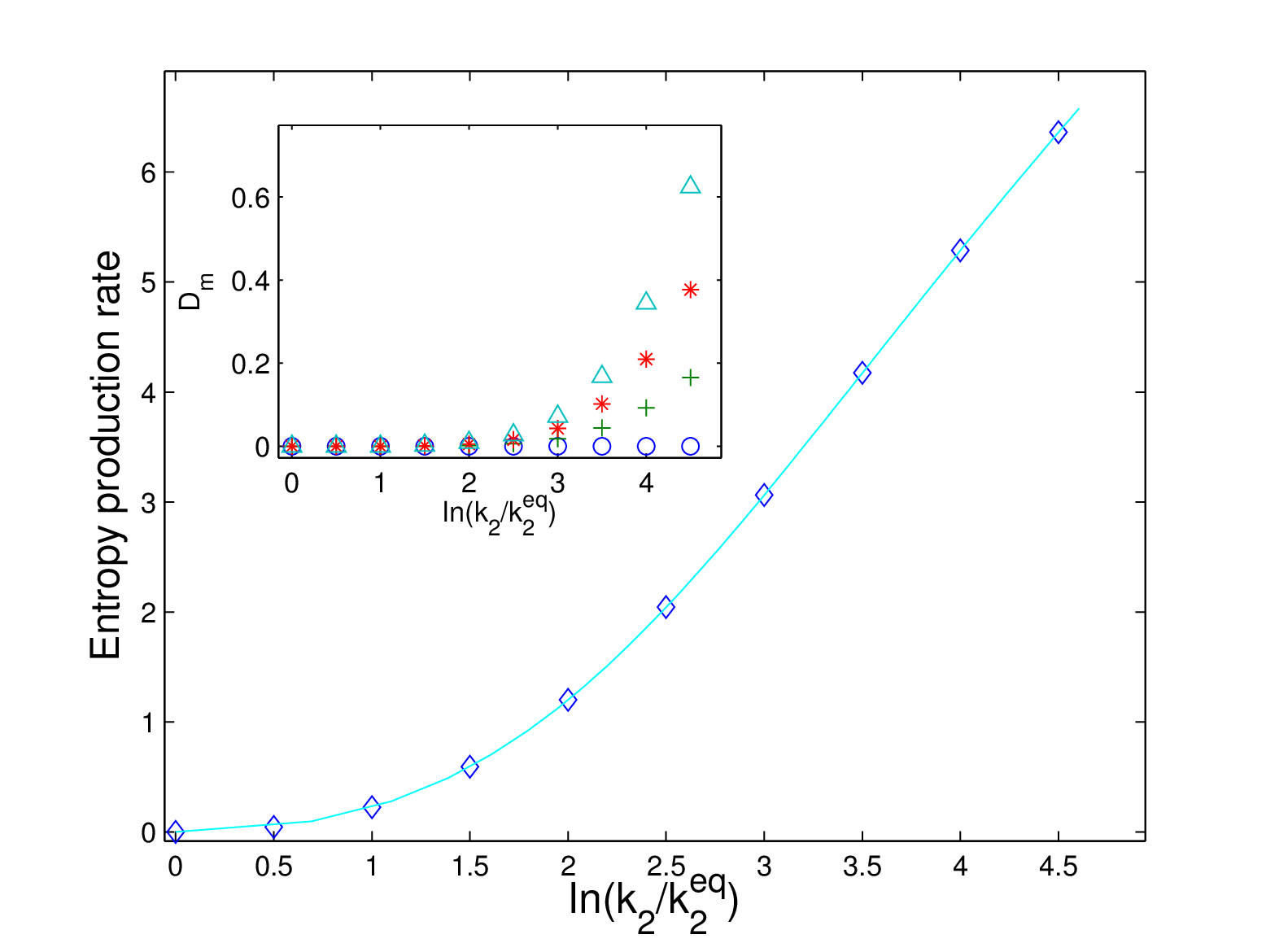}
\hspace{2mm}
\caption{\label{fig2}
 Entropy production rate (in units $s^{-1}$) as function of $\ln (k_2/k_2^{eq})$, as estimated
from $D_2$ (diamonds) using full information together with the analytical expression from Eq.~\ref{DS linear pb} (solid line).
The total number of particles and trajectory length is the same as in figure \ref{fig1} while the rates
(in units $s^{-1}$) are $k_1=1.2, k_{-1}=0.9, k_{-2}=0.4, k_{3}=0.5$, and $k_{-3}=0.2$.
In the inset, $D_2$ (circles), which is zero, $D_3$ (plus), $D_4$ (stars) and $D_5$ (triangles) are shown as
function of the same log-ratio when only partial trajectory information is used.}
\end{figure}

In fig.~\ref{fig1}, we show these estimators as a function of $k_1$,
keeping all the other rates fixed. As mentioned above, the thermodynamic driving force in this model is the affinity
$A=\ln(k_1 k_2 k_3/  k_{-1} k_{-2} k_{-3})$, which can be written  as $\ln(k_1/k_1^{eq})$
where $k_1^{eq}$ represents the equilibrium value of $k_1$.
Similarly in fig.~\ref{fig2} where $k_2$ is varied instead of $k_1$, the same driving force
is written as $\ln(k_2/k_2^{eq})$. The exact entropy production rate grows monotonously as function of this driving force
in both cases. However, the estimators vary monotonously when $k_2$ is varied, but do not when $k_1$ is varied
as shown in the inset of the figures \ref{fig1} and \ref{fig2}.

\begin{figure}[t!]
\includegraphics[scale=0.5]{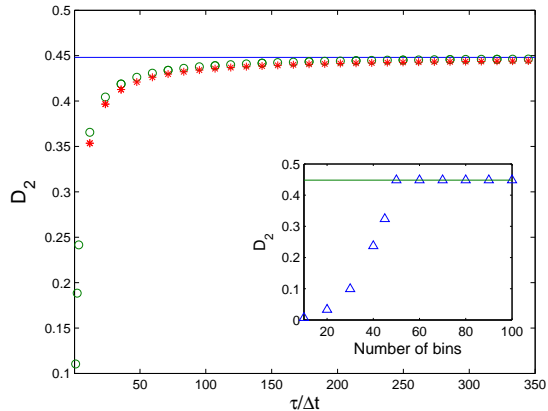}
\hspace{2mm}
\caption{\label{fig3}
Relative entropy $D_2$ as function of the ratio $\tau/\Delta t$, where $\Delta t$ represents the
duration over which the data is coarse-grained in time.
Circles correspond to a system with $50$ particles while stars correspond to a system with $5$ particles.
In the inset, the estimate of entropy production rate is shown
as function of the number of bins used to coarse-grain the particle numbers in a system of $N=50$ particles.
In both figures, the solid line represent the expected value of relative entropy.}
\end{figure}

For the case (b) and (c), we consider again trajectories with full information of the form $(n_A(t),n_B(t))$ but we assume
a finite resolution, either in particle number or in time. For case (b), we bin the particle numbers into a variable number of bins,
which affects both the relative entropy and the characteristic time $\tau$.
As seen in the inset of fig~\ref{fig3}, the estimate of the dissipation rate using coarse-grained trajectories is
 smaller than the exact value when there are less bins than particles while the exact
 value is recovered when there are more bins than particles. We observe a sharp transition between both regimes.
In case (c), we use trajectories sampled at a finite resolution or frequency $1/\Delta t$ and we keep only one
 point of the trajectory within a bin of size $\Delta t$.
As seen in fig~\ref{fig3}, this coarse-graining leads to a reduction
of the estimated entropy production except in the limit of high sampling frequency
where the exact expected value is recovered. We also note that
simulations with different $N$ can be rescaled, indicating that in this example only the
ratio of the characteristic jump time $\tau$ to the sampling time $\Delta t$ matters.
This reduction of the entropy production varies smoothly with the degree of coarse-graining,
in contrast to the case of coarse-graining via decimation over fast states where the network topology matters 
\cite{Puglisi2010_vol2010}.

\section{Applications to non-linear chemical reactions}
Finally, we discuss the estimation of dissipation from fluctuations in non-linear chemical reactions.
We start with the well-known example of
 Schl\"ogl's trimolecular reaction \cite{Schlogl1972,Gaspard2004_vol120,Vellela2009}:
\beq
\ce{A <=>[k_1][k_{-1}] X}, \,\,\,\, \ce{3X <=>[k_2][k_{-2}] 2X + B},
\eeq
where $A$ and $B$ represent two chemostats. We denote by $n_X$ the particle number of the chemical species $X$.
We recall that the corresponding concentration $[X]$ is $<n_X>/\Omega$ in terms of the extensivity parameter $\Omega$. 
In a certain range of parameters, the macroscopic equations for $[X]$ exhibit bi-stability, which means that two solutions exist for $[X]$. In contrast, at a stochastic level, the concentration interpolates between low and high concentrations.
This model is characterized by the following transition rates
\bea
w^A(n_X,n_X+1) &=& k_1 [A] \Omega, \nonumber \\
w^A(n_X,n_X-1) &=& k_{-1} n_X, \nonumber \\
w^B(n_X,n_X+1) &=& k_{-2} [B] n_X (n_X-1)/ \Omega, \\
w^B(n_X,n_X-1) &=& k_2 n_X (n_X-1) (n_X-2)/ \Omega^2,
\eea
where the superscript on the transition rates indicates which chemostat is involved in the transition,
and $[A]$ and $[B]$ are the concentration of chemostats $A$ and $B$.

When no knowledge of the chemostats and of their interaction with the system of interest is assumed,
we expect again $D_2=0$ because $D_2$ is sensitive to the average current of the variable $n_X$ which should be zero.
Intuitively, one can understand this from the following argument: if you exchange the labels of the chemostats $A$ and $B$, the net average current of $n_X$ due to interaction with $A$ and $B$ will be reversed, which means that such a current must be zero if $A$ and $B$ are not identified.
To see this more precisely, one can consider the coarse-grained
dynamics of the variable $n_X$ which is characterized by the following lumped transition rates \cite{Esposito2012_vol85}
\beq
w(n_X,n_X \pm 1)=\sum_{\nu=A,B} \frac{w^\nu(n_X,n_X \pm 1) P_{st}^\nu(n_X)}{P_{st}(n_X)},
\eeq
where $P_{st}^\nu(n_X)$ is the stationary distribution of the variable $n_X$ for the mechanism $\nu=A,B$.
In this stationary state, these distributions obey a detailed balance condition for each value of $\nu$ separately:
\beq
P_{st}^A(n_X) w^A(n_X,n_X-1) - P_{st}^A(n_X-1) w^A(n_X-1,n_X)=0,
\eeq
and
\beq
P_{st}^B(n_X) w^B(n_X,n_X-1) - P_{st}^B(n_X-1) w^B(n_X-1,n_X)=0.
\eeq
Using the equations above, it is then a simple matter to check that the local current
defined by
$J(n_X \rightarrow n_X-1) = P_{st}(n_X) w(n_X,n_X-1)  - P_{st}(n_X-1) w(n_X-1,n_X)$ is zero.
As a result $D_2$ is zero, because the relation between this local current and $D_2$ is similar to that of Eq.~\ref{D2}.
Alternatively, one can also use here the argument of probability conservation and stationarity to show that
the local current must be global, and then it must be zero due to stationarity.

\begin{figure}[t!]
\includegraphics[scale=0.5]{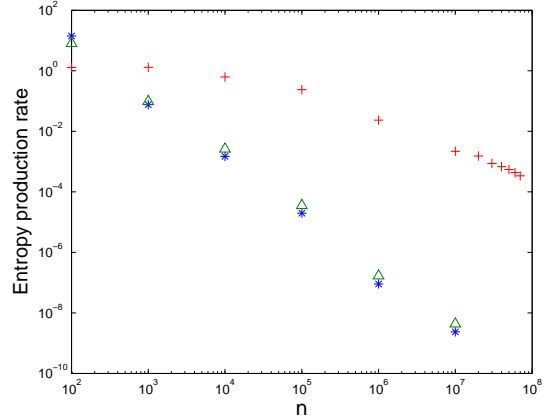}
\hspace{2mm}
\caption{\label{fig4}
Estimate of EPR using $D_2$ (stars), $D_3$ (triangles) and $D_4$ (plus) in the
Schl\"ogl's model as function of the length of the trajectory $n$.
The parameters in that figure are
$k_1 [A]=0.5$, $k_{-1}=2$, $k_2=1$,$k_{-2}=1$,
the value of the concentration of the $B$ chemostat is $[B]=3$ and
the extensivity parameter is $\Omega=10$.}
\end{figure}

In contrast with the example of the three state enzymatic model studied earlier, the dynamics of this model is Markovian
 and at equilibrium while the dynamics of the three state enzymatic model was non-Markovian and non-equilibrium in the
case of partial information.
This difference of behavior may at first appear surprising but it is not when realizing 
that the chemostats in the Schl\"ogl model are ideal. This means that
they contain an infinite number of particles. If instead they would contain a finite number of particles, the interaction between the system and the chemostats
would leave a memory that the transition has occurred. This would then imply a non-Markovian evolution as in the case of the three state enzymatic model.
In the end, in the present case, the Markovian nature of the dynamics together with the 
fact that $D_2=0$ implies that for any $m$, $D_m=0$. This is indeed
what we confirm numerically when the length of the trajectory goes to infinity as shown in the figure \ref{fig4}. Thus, at this level of description, the system is in equilibrium,
 while it would not be if the transitions involving the chemostats were properly identified, unless of course
detailed balance holds, which occurs at the concentration $[X]_{eq}=k_1 [A]/k_{-1}=k_{-2}[B]_{eq}/k_2$.
This is an example, where the lack of identification of the chemostats represents a form of
coarse-graining which has a dramatic impact in this 1D model since it prevents the identification of the NESS. 
The same point has been nicely illustrated before using linear Langevin equations \cite{VandenBroeck2010_vol82}. 
In this work, a Langevin equation is considered, which describes the dynamics of one particle in contact with two 
different thermostats. There are two friction coefficients and two white noises of different amplitudes. 
It is straightforward to show that there is an effective Langevin equation for this problem with equilibrium dynamics. 
Therefore, such a model is also a non-equilibrium one but there is no way to see it at the coarse-grained level.

This failure to identify the NESS does not occur however when the model includes at least
two variables even when chemostats are still unidentified.
In some sense, it is a topological issue, related to the Schnakenberg construction for NESS, according to which the entropy production
in a NESS can be decomposed into cycles \cite{Schnakenberg1976}.
In a 1D model even non-linear, there is no possibility to construct such a cycle, hence the entropy production is 
zero and the model is at equilibrium.
The only possibility to create a NESS in a 1D Markovian model is to use periodic potentials as discussed extensively 
in the literature on ratchet models \cite{Julicher1997_vol69,Lacoste2011_vol60}.
The situation in 2D is very different in that respect.
To illustrate this point, let us look at a 2D model, namely
the reversible Schnakenberg model \cite{Schnakenberg1979}, which may be viewed as a variant of the Brusselator model
 \cite{Nicolis1977_vol}, studied more recently in Refs~\cite{Qian2002,Xu2012}:

\beq
\ce{X <=>[k_1][k_{-1}] A}, \,\,\,\, \ce{B <=>[k_2][k_{-2}] Y}, \,\,\,\, \ce{2X + Y <=>[k_3][k_{-3}] 3X},
\eeq
where $A$ and $B$ are chemical species present in chemostats, and $X$ and $Y$ are species
with fluctuating particle numbers $n_X$ and $n_Y$.
This model presents either a monostable phase or a limit cycle depending
on the chemical potential difference
between $A$ and $B$.
 When a chemical potential difference is present, we find that one can detect the dissipation already at the level of $D_2$, because a stationary current exists in the variables $n_X$ and $n_Y$. This stationary current produces a circulation in the plane of $(n_X,n_Y)$, 
in other words, a limit cycle \cite{Qian2002,Xu2012}. In fact, a similar effect is present 
already with two linearly coupled Langevin equation where the two degrees of freedom are coupled 
to two thermostats at different temperatures. For such a model, 
the reduction from a two variables description to a one variable description \cite{Villamaina2009_vol} 
and the conditions for obtaining a limit cycle \cite{Dotsenko2013} have been studied analytically recently.

For the reversible Schnakenberg model, we can compare the estimate of dissipation using the method presented in this paper,
with that obtained using the Lebowitz-Spohn functional. Assuming the transition rates are given, the mean
entropy production rate can be obtained from this functional \cite{Lebowitz1999_vol95}, which is defined as 
 \beq
 Z(t)=\int_0^t dt' \sum_n \delta(t-t_n) \ln \frac{w(c_n,c_{n-1})}{w(c_{n-1},c_{n})},
\label{LS functional}
 \eeq
 where $t_n$ is the time of transition from state $c_{n-1}$ to $c_n$.
\begin{figure}[t!]
\includegraphics[scale=0.5]{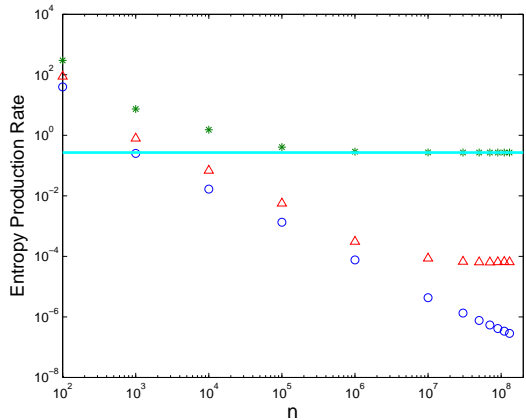}
\caption{\label{fig5}
Estimate of EPR in the non-linear
 Schnakenberg model as function of the trajectory length $n$ for various estimators (symbols)
together with the exact value calculated using Eq.~\ref{LS functional} (solid lines) for the same trajectories.
The exact expected value of the EPR is recovered from $D_2$ using full information (stars).
With partial information, one finds that $D_2$ (circles) goes to zero as $n$ increases while other estimators
such as $D_3$ (triangles) go to a finite value.
The parameters are $[A]=0.2$, $[B]=0.1$ and $\Omega=10$.
The rates (in units $s^{-1}$) are $k_1=1.5$, $k_{-1}=1$, $k_{2}=1.2$, $k_{-2}=0.2$, $k_{3}=1$ and $k_{-3}=0.3$.}
\end{figure}
 This quantity can be evaluated on the same trajectory used to determine the dissipation with $D_2$,
 and one obtains from it the EPR by evaluating $\lim_{t \rightarrow \infty}  \frac{1}{t} \langle Z(t) \rangle$.
 As shown in figure \ref{fig4}, the agreement
is very good confirming that the method is able to recover the expected value of the dissipation
 in this non-linear example.
Finally, as we did with the three states enzymatic model and with Schl\"ogl model,
it is natural to ask how this model performs if only partial information is available.

As we argued before, the specificity of the results obtained for the
Schl\"ogl model is related to the 1D nature of the model and the ideality of the reservoirs to which the observable 
of interest is coupled.
In view of this, we should expect a behavior closer to
the three states enzymatic model for the Schnakenberg model.
This is indeed what we find. In figure \ref{fig5}, one can see that
$D_2$ tends to zero when partial information is used
while $D_3$ reaches a non-vanishing plateau at large $n$, just as in the three states enzymatic model with partial information.
This shows that for this non-linear example, our method is still
able to distinguish equilibrium from non-equilibrium fluctuations. It is remarkable that
this can be done using only partial information of one chemical species.

\section{Conclusions}

In this paper, we have illustrated a general method which is able to
infer the amount of dissipation present in the fluctuations of
a chemical system.
There are no specific limitations to the complexity of the chemical system, which can involve an arbitrary number of linear or non-linear reactions; for simplicity, we have illustrated the principles of the method with the Schl\"ogl and Schnakenberg models.
Even more remarkable is the fact that the method is non-invasive and does not require any knowledge of the underlying dynamics, 
which makes it ideally suited for applications in chemistry or biology. Since fluctuations of a chemical system can be viewed as a 
form of noise, one can say that the method is able to identify the nature of the noise, or at least 
to extract some relevant information contained in the noise.

The method requires in principle a detailed information about the fluctuations, which may be difficult to obtain in practice. 
To address this issue, we have studied the robustness with respect to a reduction in the amount of 
information present in the input data, a process which we call coarse-graining. We have considered the effect of reducing 
the number of recorded chemical species and the effect of a limited resolution either in particle number or in time. 
We have shown that in such situations, the method is still able to provide useful information. 
In particular, it can systematically distinguish equilibrium fluctuations from non-equilibrium ones, even when 
traditionally methods fail, as shown in the example of the three-state enzymatic network.
When applied to biological systems, it could serve as a means of investigation of active biological systems, 
to be used in connection with micro-rheology techniques \cite{Mizuno2007_vol315}.
In principle, the method could provide more than just a yes/no answer to the question, to whether the system is in 
equilibrium, since it can provide an estimate of the dissipation. However, it is likely that accurate estimates will 
be more difficult to obtain than a yes/no answer, since the quality of the estimation depends crucially
on the quantity and on the quality of the input data.

In this paper, we have considered various dynamics involving a finite number of states as in the work of Roldan et al. \cite{Roldan2010_vol105}.
It would be interesting to apply this method to continuous data sets, which would make
closer connections to the original work of P. Gaspard \cite{Gaspard2004_vol117a} and would be  
useful for many applications to biological or chemical systems. Such an extension has already been
achieved in an experiment using manipulated colloids which can be described by a linear Langevin equation \cite{Andrieux2008a}.
It remains to be seen whether this method can be exploited to study more complex experimental systems such as biological ones.

These applications ideally should involve a small chemical or biochemical system 
in which the fluctuations of concentration of some chemical species can be measured with a high temporal resolution. 
The technique of fluorescence correlation spectroscopy is a possible candidate for experiments of this kind 
since it allows to perform measurements in a small volume,
which is nowadays easily done in a microfluidic device, with the possibility of a good temporal resolution at 
the single molecule level.
We hope that our work could motivate experiments along this line. 

\section{Acknowledgments}
We acknowledge stimulating discussions with A. Lemarchand, H. Berthoumieux, J. M. R. Parrondo and L. Jullien.

%

\end{document}